# Extinction debt repayment via timely habitat restoration

Katherine Meyer[1]

**Abstract**

Habitat destruction threatens the viability of many populations, but its full consequences can take considerable time to unfold. Much of the discourse surrounding extinction debts—the number of species that persist transiently following habitat loss, despite being headed for extinction—frames ultimate population crashes as the means of settling the debt. However, slow population decline also opens an opportunity to repay the debt by restoring habitat. The timing necessary for such habitat restoration to rescue a population from extinction has not been well studied. Here we determine habitat restoration deadlines for a spatially implicit Levins/Tilman population model modified by an Allee effect. We find that conditions that hinder detection of an extinction debt also provide forgiving restoration timeframes. Our results highlight the importance of transient dynamics in restoration and suggest the beginnings of an analytic theory to understand outcomes of temporary press perturbations in a broad class of ecological systems.

*Elements:* Main manuscript, Appendix A.

## Introduction

Biodiversity underpins many ecosystem services [Balvanera et al. 2006], and continues to decline on a global scale [Butchart et al. 2010]. A key culprit is human modification of species' habitats, from expansion of agricultural land use to sea floor trawling [Millenium Ecosystem Assessment]. However, loss of species does not necessarily occur immediately after a habitat loss. Tilman and colleagues coined the term "extinction debt" to describe the number of species that transiently persist following habitat destruction, despite being deterministically headed to extinction [Tilman et al. 1994]. While much of the discourse surrounding extinction debts has framed delayed extinction as the ultimate means of settling the debt [e.g. Kolk and Naaf, 2015], other studies have highlighted opportunities and strategies for repaying the debt by habitat restoration [Huxel and Hastings 1999, Hanski 2000, Wearn et al. 2012]. Hanski's (2000) model of Finnish forest species included a restoration strategy that saved species if implemented immediately but lost them if delayed by 30 years. This result highlights the time-sensitivity of habitat restoration for recovering species from an extinction debt. The question of just how soon habitat must be restored to bring species out of extinction debts remains largely unexplored. This paper examines the timeframes necessary for effective habitat restoration using a conceptual, spatially implicit population model adapted from [Tilman et al. 1994], [Levins and Culver 1971], and [Chen and Hui 2009].

[1]University of Minnesota Department of Mathematics, Minneapolis, MN. Email: `meye2098@umn.edu`

The primary difference between the model employed here and the Tilman (1994) model is incorporation of Allee effects, which in many real-world cases may prevent a population from recovering from low abundances despite habitat restoration. For example, sexually-reproducing sessile organisms may face lower colonization rates at low abundances because their gametes collocate infrequently [Hastings and Gross 2012], while pack-forming animals with few packs may not generate enough dispersers to successfully form new groups [Courchamp and Gascoigne 2008]. Such reductions in colonization efficiency at low populations can introduce Allee thresholds, and in this context, repaying an extinction debt requires restoring habitat not just before extinction but before the population drops below the threshold. Below we compute extinction debt repayment deadlines for a modified single-species Tilman/Levins population model that includes a colonization component Allee effect. We determine how initial, transient, and final habitat destruction levels impact the timeframe for habitat restoration, and connect the problem of extinction debt repayment to a rich set of questions about press perturbation intensity and duration.

**Model**

*Overview*. The present population model is adapted from the single-species case of the spatially implicit Tilman/Levins population model [Tilman et al. 1994, Levins and Culver 1971],

$$\frac{dp}{dt} = cp(1 - D - p) - mp \qquad (1)$$

where $p$ represents the proportion of sites occupied by individuals in a grid-like habitat, $m$ is the mortality rate, $c$ is the colonization efficiency, and $D$ is the proportion of habitat sites destroyed.

We incorporate an Allee effect and time-varying habitat destruction into this model as follows:

$$\frac{dp}{dt} = C(p) \cdot p \cdot (1 - D(t) - p) - mp \qquad (2a)$$

$$\text{where} \quad C(p) = c\frac{p}{a+p} \qquad (2b)$$

represents a colonization efficiency that varies with $p$ and is responsible for the Allee effect. The functional form of $C(p)$ fits the more general form $c\frac{p-bA}{aA+p}$ used by [Chen and Hui 2009] to represent colonization component Allee effects, with parameters $b=0$ and $A=1$. In formulation (2b), colonization efficiency is nonnegative and increases in a saturating manner with half-saturation constant $a$. The function $D(t)$ gives the time-dependent proportion of habitat destroyed.

A metapopulation interpretation of (1) is also possible, in which $p$ represents proportion of habitat patches occupied by populations, and $m$ is a local extinction rate. For brevity we will use population terminology; however, results could also be interpreted in a metapopulation context, with "Allee effect" replaced with "Allee-like effect" as in [Hastings and Gross 2012].

*Constant habitat destruction*. Figure 1A summarizes the equilibria and bifurcation structure of this system when $D(t)$ is a constant $D$ between 0 and 1, representing the proportion of habitat unavailable for colonization. For any $D$, $p = 0$ is a stable equilibrium.

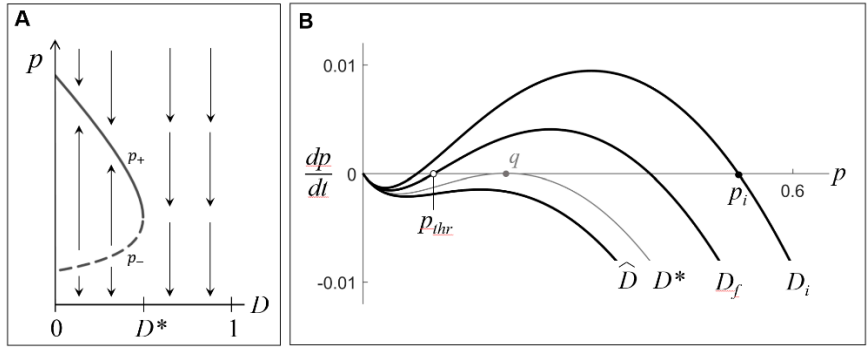

**Figure 1** (**A**) Bifurcation diagram for spatially implicit population model (2) with respect to habitat destruction parameter D. Solid and dashed lines show stable and unstable branches of equilibria, respectively; arrows indicate direction of population change. (**B**) Rates of change in abundance as a function of abundance for representative values of habitat destruction.

If $D$ is less than a critical level of habitat destruction

$$D^* = 1 - m/c - 2\sqrt{ma/c},$$

there exist additional stable (solid line, $p_+$) and unstable (dashed line, $p_-$) equilibria at

$$p_+, p_- = \frac{\left(1 - D - \frac{m}{c}\right) \pm \sqrt{\left(1 - D - \frac{m}{c}\right)^2 - \frac{4ma}{c}}}{2}$$

To ensure these equilibria exist and are positive for some positive values of $D$, we assume that $0 < m < c$ (maximum colonization rate exceeds mortality rate) and $0 < a < \left(1 - \frac{m}{c}\right)^2 / \frac{4m}{c}$. The stable positive equilibrium $p_+$ represents the long-term abundance of a viable population, while the unstable equilibrium $p_-$ represents an Allee threshold. At $D = D^*$ these positive equilibria coalesce in a saddle-node bifurcation, and for $D > D^*$ they cease to exist, so any population decreases towards zero. A positive population $p$ when $D > D^*$ represents an extinction debt, and the population will crash if no restoration occurs.

***Variable habitat destruction***. The following piecewise-constant habitat destruction parameter represents habitat destruction followed by restoration:

$$D(t) = \begin{cases} D_i, & t < 0 \\ \widehat{D}, & 0 \leq t \leq T \\ D_f, & T < t. \end{cases} \quad (2c)$$

The scenarios of interest start with enough habitat to support a population ($D_i < D^*$), introduce an extinction debt by increasing habitat destruction to $\widehat{D} > D^*$, then restore habitat at time $T$ to achieve a final value $D_f < D^*$. The question at hand is how quickly the habitat restoration must occur to allow recovery of the population to a positive stable equilibrium.

The dynamics that determine the answer are best conveyed via an example; we consider parameter values $c=0.25$, $a=0.1$, $m=0.1$, $D_i=0$, $\widehat{D}=0.25$, and $D_f=0.1$. For this set of parameters, the critical habitat destruction value is $D^* = 0.2$. Figure 1B shows $dp/dt$ as a function of $p$ for $D_i$, $\widehat{D}$, $D_f$, and $D^*$. Suppose the population starts at the positive stable equilibrium $p_i$ corresponding to $D_i$. At time $t = 0$, habitat destruction increases to $\widehat{D}$, introducing an extinction debt. As time passes, $p$ decreases toward 0 until at time $T$ habitat is restored, changing $D$ to $D_f$. At this moment, the position of $p$ relative to the unstable equilibrium $p_{thr}$ associated with $D_f$ determines the long-term fate of the population. Figure 2 shows sample trajectories for $T=100$ (dashed line), $T=122$ (dot-dashed line), and $T=140$ (dotted line). For $T$ sufficiently small (dashed), $p(T)$ exceeds $p_{thr}$ and the population recovers; for $T$ large enough (dotted), $p(T)$ is below $p_{thr}$ and the population crashes. The boundary between these cases occurs when $p(T)$ is exactly $p_{thr}$ (dot-dash). For the parameter choices of Figures 1B and 2, the boundary between recovery and extinction occurs at $T^*=122$.

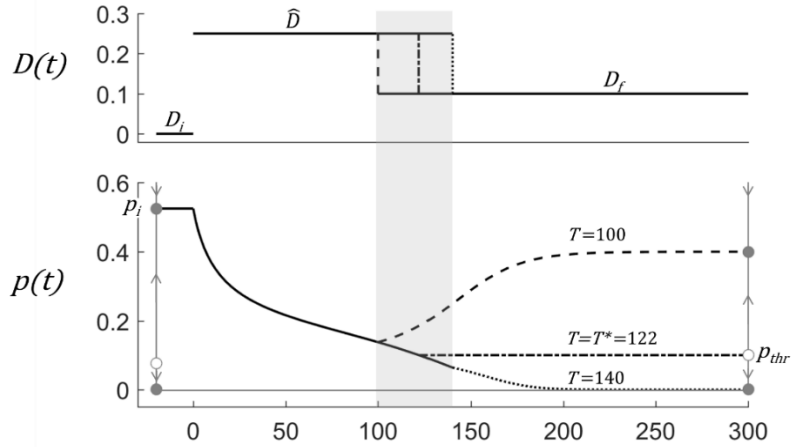

**Figure 2** Outcomes of three durations of low habitat availability: short (dash), medium (dash-dot), and long (dot). (**top**) Habitat destruction levels as a function of time. (**bottom**) Corresponding population trajectories. Grey vertical lines at $t = -25$ and $t = 300$ indicate dynamics under initial and final habitat destruction levels. Filled circles are stable equilibria, open circles are unstable.

For general model parameters, separation of variables applied to equation (2a) gives the critical time $T^*$ such that $p(T^*)=p_{thr}$:

$$\int_{p_i(D_i)}^{p_{thr}(D_f)} \frac{dp}{C(p) \cdot p \cdot (1-\widehat{D}-p) - mp} = \int_0^{T^*} dt = T^* \qquad (3)$$

Note that instead of calculating a population change in terms of time, equation (3) solves for the time $T^*$ that it takes the population to decrease from $p_i$ to $p_{thr}$. $T^*$ represents the extinction debt repayment deadline, and depends on initial habitat destruction $D_i$ (which determines initial abundance $p_i$), transient habitat destruction $\widehat{D}$ (which affects the rate of population decline) and final habitat destruction $D_f$ (which determines the final Allee threshold $p_{thr}$).

The software MATLAB R2016b (The MathWorks, Inc) was used to calculate the extinction debt repayment deadlines $T^*$ according to equation (3) for a range of $D_i$, $\widehat{D}$, and $D_f$ values. Using the parameters of Figure 1B as baseline values, Figure 3 shows $T^*$ as a function of $\widehat{D}$ (Figure 3A), $D_i$ (Figure 3B), and $D_f$ (Figure 3C).

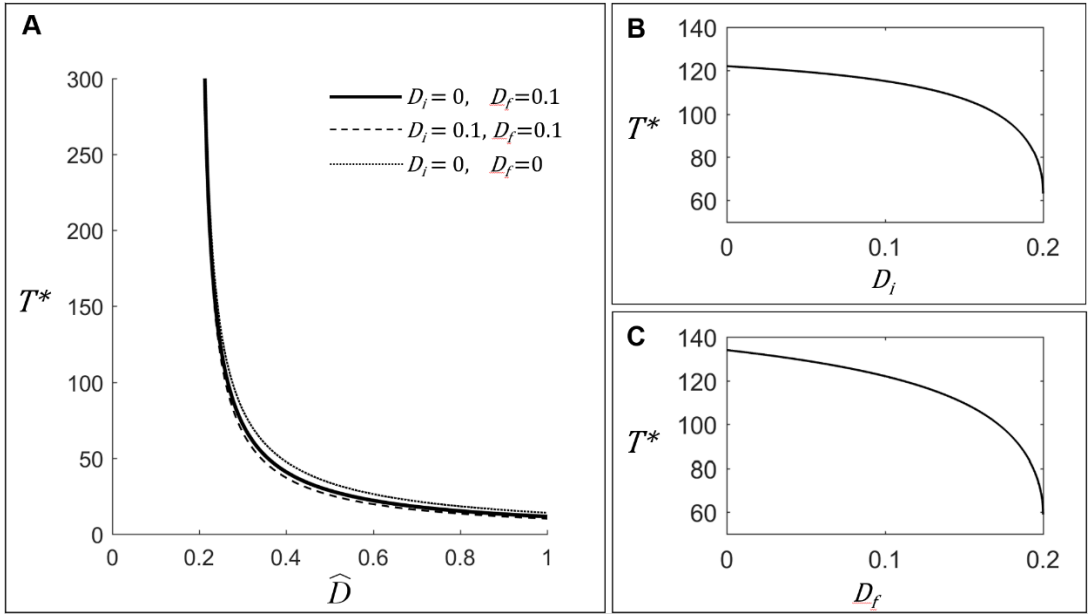

**Figure 3** Habitat restoration deadlines as a function of temporary (**A**), initial (**B**), and final (**C**) levels of habitat destruction. Baseline parameters are the same as those for Figure 1B.

**Discussion**

To avoid extinction of an Allee population threatened by habitat destruction, habitat must be restored not just before extinction occurs but before an Allee threshold is crossed. In practice, the data needed to accurately parameterize an Allee effect may be infeasible to obtain for a species of interest, and parameter estimation could significantly alter the quantitative predictions of the model presented here [Labrum 2001]. However, even qualitative conclusions might be useful for conservation management. The relationships described below, evident in Figure 3, hold across parameter values such that $0 < m < c$, $0 < a < \left(1 - \frac{m}{c}\right)^2 / \frac{4m}{c}$, and $\widehat{D} > D^* > D_i, D_f > 0$, provided that $p_{thr} < p_i$ (Online Appendix A).

*Deadline dependence on temporary habitat destruction*. As might be expected, larger temporary habitat destruction levels ($\widehat{D}$) create shorter deadlines for habitat restoration ($dT^*/d\widehat{D} < 0$, Figure 3A). This relationship stems from the magnification of the negative growth rate $dp/dt$ as $\widehat{D}$ increases beyond $D^*$: more severe habitat destruction causes the population to decrease more quickly toward the Allee threshold between times 0 and $T$. On the other hand, as the temporary habitat destruction $\widehat{D}$ decreases toward $D^*$, the minimum level needed to introduce an extinction debt, the negative growth rate weakens toward zero at the degenerate equilibrium $q$ of $D=D^*$ (Figure 1B). Consequently, the habitat restoration deadline $T^*$ grows without bound as $\widehat{D}$ decreases to $D^*$, giving the vertical asymptote of Figure 3A. The deadline for habitat restoration is therefore quite sensitive to small changes in $\widehat{D}$ near $D^*$, and habitat destruction just over the critical value $D^*$ may allow very long deadlines for habitat restoration. This outcome relates closely to Hanski and Ovaskainen's prediction that species just past an extinction threshold exhibit a particularly large time delay to extinction [Hanski 2002]. Though Hanski and Ovaskainen were concerned with time to extinction rather than to an Allee threshold, the slowness of transient dynamics near a bifurcation underlies both phenomena. It is perhaps encouraging that the same conditions that obfuscate detection of crossing a critical level of habitat destruction also give the most forgiving habitat restoration deadlines.

*Deadline dependence on initial habitat destruction*. As Figure 3B illustrates, systems that begin with greater initial habitat destruction ($D_i$) have shorter deadlines for habitat restoration ($dT^*/dD_i < 0$). This relationship stems from the influence of habitat destruction $D_i$ on initial population level $p_i$. As $D_i$ increases to the critical level $D^*$, the initial (equilibrium) population $p_i$ decreases, approaching the saddle-node bifurcation as in Figure 1A. In turn, a lower initial population reduces the time needed to reach the Allee threshold while the population declines under the high habitat destruction level $\hat{D}$.

Furthermore, the sensitivity of the deadline to initial habitat destruction increases as $D_i$ approaches $D^*$ ($d^2T^*/dD_i^2 < 0$), with the slope of the relationship ($dT^*/dD_i$) becoming arbitrarily large in magnitude near $D^*$. This nonlinearity means that the impact of initial habitat destruction levels on the deadline for restoration changes only modestly over a range of values (see the slight change in $T^*$ for $D_i = 0$ vs. $D_i = 0.1$ in Figure 3A), but becomes larger near the critical level of habitat destruction, $D^*$.

The dependence of deadline on initial habitat destruction through the initial population abundance suggests an additional management strategy: in situations where the habitat restoration deadline cannot be met but conservation is a high priority, artificially increasing the abundance of the species could extend the deadline by placing the population further from the Allee threshold, making it take longer to cross. Without habitat restoration this would simply delay extinction, but in concert with restoration it could support long-term persistence.

*Deadline dependence on final habitat destruction*. Restoring additional habitat buys extra time for restoration ($dT^*/dD_f < 0$, Figure 3C). This is because reducing $D_f$ lowers the final Allee threshold $p_{thr}$, allowing more time to pass during the declining phase ($D=\hat{D}$) before this threshold is crossed. As with the relationship between deadline and initial habitat destruction, the sensitivity of deadline to final habitat destruction $D_f$ is high when $D_f$ is close to the critical habitat destruction value $D^*$; in particular, $dT^*/dD_f$ decreases without bound as $D_f$ approaches $D^*$. Therefore, restoring an additional percentage of habitat extends the deadline most when final habitat destruction is just below the critical level.

*Future Directions*. The restoration deadline question considered here represents a larger class of problems involving the reversibility of temporary environmental changes, which arise both within ecology and beyond. Hughes and colleagues [2013] pointed out that slow ecological regime shifts afford a window of opportunity for corrective action, and called for further modelling research to understand the non-equilibrium dynamics and timeframes of these windows. Ratajczak and colleagues [2017] simulated how both intensity and duration of ecological press perturbations such as grazing, nutrient loading, and fire suppression influence outcomes. The general pattern that they detected—a negative but saturating relationship between critical duration and press intensity—matches our results (Figure 3A). Transient press perturbations are also relevant in fields beyond ecology: heightened North Atlantic meltwater flux of sufficient duration and intensity is thought to have triggered historic switches in ocean circulation pattern [Cessi 1994], while elevated greenhouse gas forcing may tip the Artic sea to an ice-free state that persists even after reduction in forcing [Eisenman and Wettlaufer 2009].

The separation-of-variables technique used in equation (3) to calculate habitat destruction deadlines can model the outcome of any piecewise-constant parameter change in a system of a single variable. Indeed, Cesssi [1994] used the same technique to calculate combinations of meltwater duration and intensity sufficient to trigger a shift in ocean circulation. However, many interesting questions don't yield to separation-of-variables. For example, one might wish to calculate extinction debt repayment deadlines in a full multispecies hierarchical competition model [Tilman et al. 1994, Chen and Hui 2009], in the context of a continuous change in habitat availability (due to gradual restoration, for example), or in spatially explicit models [Huxel and Hastings 1999, Hanski 2000]. Numerical simulations could certainly be used to tackle these questions on a case-by-case basis. On the other hand, analytical treatments of higher dimensional systems and continuous parameter changes are also possible [Ritchie et al. 2017]. Further mathematical work on transient parameter changes could bring broader insights into the dynamics they induce.

**Conclusion**

This study has highlighted the temporal aspect of repaying extinction debts via habitat restoration, but is not meant to suggest that restoring habitat before an appropriate deadline eliminates the damage of the original habitat destruction. Indeed, recovering landscapes don't always reach pre-disturbance levels of organismal abundance, species richness or geochemical cycling, and even when they do, deficits of ecosystem services during recovery accrue a "recovery debt" [Moreno-Mateos et al. 2017]. With the understanding that restoration of degraded systems complements protection of pristine systems [Possingham et al. 2015], the model presented here makes the optimistic point that populations in an extinction debt induced by habitat loss needn't be doomed if habitat is restored in a timely manner. The precise meaning of "timely" depends on multiple factors, with deadlines for habitat restoration becoming tighter when initial or restored habitat is close to the brink at which a population is no longer viable, or when the temporary habitat loss is severe.

Extinction debt repayment represents one instance of potential recovery from a transient environmental change. Given humans' widespread impacts on Earth systems and growing efforts to mitigate these impacts, many more such instances are expected. Developing further mathematical techniques to describe outcomes in multi-dimensional (e.g. multi-species) and non-autonomous (e.g. continuous parameter change) systems will improve our ability to predict not only habitat restoration outcomes, but also repercussions of transient changes in a broad class of systems.

**Acknowledgements**

This work was supported by an NSF Graduate Research Fellowship to Katherine Meyer (grant number 00039202). Thanks go to David Tilman, who introduced the author to the idea of an extinction debt and provided feedback on an initial draft, to Forest Isbell, who encouraged dissemination of these results, Richard McGehee, who consulted on mathematics, and to Allison Shaw, Lauren Sullivan, Evelyn Strombom, Morganne Igoe, and Julie Sherman, who provided feedback on the manuscript.

**Literature Cited**

Butchart, S. H. M., et al. 2010. Global biodiversity: indicators of recent declines. Science 328: 1164–1168.

Cessi, P. 1994. A simple box model of stochastically forced thermohaline flow. Journal of Physical Oceanography 24: 1911–1920.

Chen, L., and C. Hui. 2009. Habitat destruction and the extinction debt revisited: the Allee effect. Mathematical Biosciences 221: 26–32.

Courchamp, R., and J. Gascoigne. 2008. Allee Effects in Ecology and Conservation. Oxford: Oxford University Press.

Eisenman, I. and J. S. Wettlaufer. 2009. Nonlinear threshold behavior during the loss of Arctic sea ice. Proceedings of the National Academy of Sciences 106: 28–32.

Hastings, A., and L. J. Gross. 2012. Encyclopedia of Theoretical Ecology. University of California Press, Berkeley and Los Angeles.

Hanski, I. 2000. Extinction debt and species credit in boreal forests: modeling the consequences of different approaches to biodiversity conservation. Annales Zoologici Fennici 37: 271-280.

Hanski, I., and O. Ovaskainen. 2002. Extinction debt at extinction threshold. Conservation Biology 16: 666–673.

Hughes, T. P., C. Linares, V. Dakos, I. A. van de Leemput, and E. H. van Nes. 2013. Living dangerously on borrowed time during slow, unrecognized regime shifts. Trends in Ecology & Evolution 28: 149–155.

Huxel, G. R., and A. Hastings. 1999. Habitat loss, fragmentation, and restoration. Restoration Ecology 7: 309–315.

Kolk, J., and T. Naaf. 2015. Herb layer extinction debt in highly fragmented temperate forests – Completely paid after 160 years? Biological Conservation 182: 164-172.

Labrum, M. 2001. Allee effects and extinction debt. Ecological Modeling 222: 1205–1207.

Levins, R., and D. Culver. 1971. Regional coexistence of species and competition between rare species. Proceedings of the National Academy of Sciences of the United States of America 68: 1246–1248.

Millennium Ecosystem Assessment, 2005. Ecosystems and Human Well-being: Biodiversity Synthesis. World Resources Institute, Washington, DC.

Moreno-Mateos, D., E. B. Barbier, P. C. Jones, H. P. Jones, J. Aronson, J. A. Lopez-Lopez, M. L. McCrackin, P. Meli, D. Montoya, and J. M. Rey Benayas. 2017. Anthropogenic ecosystem disturbance and the recovery debt. Nature Communications 8: 14163. doi:10.1038/ncomms14163

Possingham, H. P., M. Bode, and C. J. Klein. 2015. Optimal conservation outcomes require both restoration and protection. PLOS Biology 13: e1002052. doi: 10.1371/journal.pbio.1002052.

Ratajczak, Z., P. D'Odorico, S. L. Collins, B. T. Bestelmeyer, F. I. Isbell, and J. B. Nippert. 2017. The interactive effects of press/pulse intensity and duration on regime shifts at multiple scales. Ecological Monographs 87: 198–218.

Ritchie, P., Ö. Karabacak, and J. Sieber. 2017. Inverse-square law between time and amplitude for crossing tipping thresholds. arXiv 1709.02645v2 [math.DS].

Tilman, D., R. M. May, C. L. Lehman, and M. A. Nowak. 1994. Habitat destruction and the extinction debt. Nature 371: 65–66.

Wearn, O. R., D. C. Reuman., and R. M. Ewers. 2012. Extinction Debt and Windows of Conservation Opportunity in the Brazilian Amazon. Science 337: 228-232.

**Appendix A: Findings across Parameter Space**

Here we show that the dependences of $T^*$ on $D_i$, $\widehat{D}$, and $D_f$ discussed in the main article hold across parameter values $c > 0$, $0 < m < c$, $0 < a < \left(1 - \frac{m}{c}\right)^2 / \frac{4m}{c}$, and $\widehat{D} > D^* > D_i$, $D_f > 0$, provided that $p_{thr} < p_i$.

For brevity, denote the function giving $dp/dt$ for fixed habitat destruction $D$ as

$$f(p; D) = c \frac{p}{a+p} \cdot p \cdot (1 - D - p) - mp \tag{A1}$$

Recall from equation (3) that the timeline for habitat restoration $T^*$ depends on initial, temporary, and final habitat destruction values $D_i$, $\widehat{D}$, and $D_f$ according

$$T^* = \int_{p_i(D_i)}^{p_{thr}(D_f)} \frac{dp}{f(p;\widehat{D})} \tag{A2}$$

and to the equations for $p_+$ and $p_-$, which imply that

$$p_i(D_i) = \frac{\left(1-D_i-\frac{m}{c}\right) + \sqrt{\left(1-D_i-\frac{m}{c}\right)^2 - \frac{4ma}{c}}}{2} \tag{A3}$$

$$p_{thr}(D_f) = \frac{\left(1-D_f-\frac{m}{c}\right) - \sqrt{\left(1-D_f-\frac{m}{c}\right)^2 - \frac{4ma}{c}}}{2} \tag{A4}$$

*Relationship 1: $dT^*/d\widehat{D} < 0$*. We have from equation (A2) that

$$\frac{dT^*}{d\widehat{D}} = \frac{d}{d\widehat{D}} \int_{p_i}^{p_{thr}} \frac{1}{f(p,\widehat{D})} dp. \tag{A5}$$

Because the integrand and its partial derivative with respect to $\widehat{D}$ are continuous away from $p=0$, the Leibniz integral rule gives

$$\frac{dT^*}{d\widehat{D}} = \int_{p_i}^{p_{thr}} \frac{d}{d\widehat{D}} \left(\frac{1}{f(p,\widehat{D})}\right) dp. \tag{A6}$$

Now the integrand in equation (A6) is

$$\frac{d}{d\widehat{D}} \left(\frac{1}{f(p,\widehat{D})}\right) = \frac{-1}{(f(p,\widehat{D}))^2} \frac{df}{d\widehat{D}} \tag{A7}$$

and the first term of this product is negative. The second term is

$$\frac{df}{d\widehat{D}} = \frac{d}{d\widehat{D}} \left(\frac{c_{max}p}{a+p} \cdot p \cdot (1 - \widehat{D} - p) - mp\right) = -\frac{c_{max}p^2}{a+p},$$

and therefore negative. Hence the integrand is the product of two negative terms and is positive, but because $p_{thr} < p_i$, the integral of this positive quantity is negative. //

*Relationship 2: T\* grows without bound as $\widehat{D}$ decreases to D\**

Let $p_*$ denote the location in state space of the bifurcation at $D = D^*$; i.e. $f(p_*; D^*) = 0$ and $\left.\frac{df}{dp}\right|_{(p_*; D^*)} = 0$. Since $f(p; \widehat{D})$ is a rational function, it is continuously differentiable away from its poles. Its derivative with respect to $p$ is continuous in both $p$ and $\widehat{D}$ over a compact region $[p_*, p_i] \times [D^*, D^* + \varepsilon]$ for some $\varepsilon > 0$, and is therefore bounded over the same region. Let $C$ represent an upper bound on $\left|\frac{\partial f}{\partial p}\right|$ over $[p_*, p_i] \times [D^*, D^* + \varepsilon]$. For notational ease, let $F(p; \widehat{D}) = -f(p; \widehat{D})$. Then $F$ also has bounded partial derivative $\left|\frac{\partial F}{\partial p}\right| \leq C$ on $[p_*, p_i] \times [D^*, D^* + \varepsilon]$. The Mean Value Theorem implies that for $p_* \leq p \leq p_i$ and $D^* \leq \widehat{D} \leq D^* + \varepsilon$,

$$F(p; \widehat{D}) \leq F(p_*; \widehat{D}) + C(p - p_*). \tag{A8}$$

We also have that

$$T^* = \int_{p_i}^{p_{thr}} \frac{dp}{f(p;\widehat{D})} = \int_{p_{thr}}^{p_i} \frac{dp}{-f(p;\widehat{D})} = \int_{p_{thr}}^{p_i} \frac{dp}{F(p;\widehat{D})} \geq \int_{p_*}^{p_i} \frac{dp}{F(p;\widehat{D})}. \tag{A9}$$

with the final inequality following from non-negativity of $F$.

Inequality (A8) implies that $\frac{1}{F(p;\widehat{D})} \geq \frac{1}{F(p_*;\widehat{D}) + C(\widehat{D})(p - p_*)}$. Combining with (A9) yields

$$T^* \geq \int_{p_*}^{p_i} \frac{dp}{F(p_*; \widehat{D}) + C(p - p_*)}$$

$$= \frac{1}{C}\left[\ln\left(F(p_*; \widehat{D}) + C(p - p_*)\right)\right]_{p_*}^{p_i}$$

$$= \frac{1}{C}\left[\ln\left(F(p_*; \widehat{D}) + C(p_i - p_*)\right) - \ln\left(F(p_*; \widehat{D})\right)\right] \tag{A10}$$

In the limit as $\widehat{D}$ decreases to $D^*$, $F(p_*; \widehat{D})$ decreases to zero, $\ln\left(F(p_*; \widehat{D}) + C(p_i - p_*)\right)$ goes to the finite quantity $\ln(C(p_i - p_*))$, and $-\ln\left(F(p_*; \widehat{D})\right)$ goes to positive infinity. This implies that both the expression (A10) and $T^*$ go to positive infinity as $\widehat{D}$ decreases to $D^*$. //

*Relationship 3: dT\*/dD$_i$ < 0.* We have

$$\frac{dT^*}{dD_i} = \frac{dT^*}{dp_i}\frac{dp_i}{dD_i}. \tag{A11}$$

From equation (A2) and the Fundamental Theorem of Calculus the first term in this product is

$$\frac{dT^*}{dp_i} = -\frac{1}{f(p_i;\widehat{D})} \tag{A12}$$

and this term is positive, because $\widehat{D} > D^*$ implies $f(p_i; \widehat{D}) < 0$.

For the second term in the product, differentiation of $p_i(D_i)$ with respect to $D_i$ gives

$$\frac{dp_i}{dD_i} = -\frac{1}{2}\left(1 + \frac{1-D_i-\frac{m}{c}}{\sqrt{\left(1-D_i-\frac{m}{c}\right)^2 - \frac{4ma}{c}}}\right) \quad (A13)$$

The condition $D_i < D^*$ implies that both $1 - D_i - \frac{m}{c}$ and $\sqrt{\left(1 - D_i - \frac{m}{c}\right)^2 - \frac{4ma}{c}}$ are positive, so $dp_i/dD_i < 0$. Therefore $dT^*/dD_i$ is the product of a positive and a negative term, and is negative. //

*Relationship 4: $d^2T^*/dD_i^2 < 0$.*

$$\frac{d^2T^*}{dD_i^2} = \frac{d}{dD_i}\left[\frac{dT^*}{dD_i}\right] = \frac{d}{dD_i}\left[\frac{dT^*}{dp_i}\cdot\frac{dp_i}{dD_i}\right] = \frac{d}{dD_i}\left[\frac{dT^*}{dp_i}\right]\cdot\frac{dp_i}{dD_i} + \frac{dT^*}{dp_i}\cdot\frac{d}{dD_i}\left[\frac{dp_i}{dD_i}\right]$$

$$= \left(\frac{d}{dp_i}\left[\frac{dT^*}{dp_i}\right]\cdot\frac{dp_i}{dD_i}\right)\cdot\frac{dp_i}{dD_i} + \frac{dT^*}{dp_i}\cdot\frac{d^2p_i}{dD_i^2}$$

$$= \underbrace{\frac{d^2T^*}{dp_i^2}}_{(I)}\underbrace{\left(\frac{dp_i}{dD_i}\right)^2}_{(II)} + \underbrace{\frac{dT^*}{dp_i}}_{(III)}\cdot\underbrace{\frac{d^2p_i}{dD_i^2}}_{(IV)} \quad (A14)$$

We will show that $(I) < 0$, $(II) \geq 0$, $(III) > 0$, and $(IV) < 0$, implying (A14) is negative.

$(I)$ must be negative, since

$$\frac{d^2T^*}{dp_i^2} = \frac{d}{dp_i}\left[\frac{dT^*}{dp_i}\right] = \frac{d}{dp_i}\left[\frac{-1}{f(p_i;\widehat{D})}\right] = \frac{1}{f(p_i;\widehat{D})^2}\frac{\partial f(p_i;\widehat{D})}{\partial p_i}, \quad (A15)$$

$\frac{\partial f(p_i;D_i)}{\partial p_i} < 0$ ($p_i$ is a stable equilibrium when $D = D_i$), and $\frac{\partial f(p_i;\widehat{D})}{\partial p_i} < \frac{\partial f(p_i;D_i)}{\partial p_i}$ for $\widehat{D} > D_i$.

$(II)$ is nonnegative (a square).

$(III)$ is

$$\frac{dT^*}{dp_i} = \frac{-1}{f(p_i;\widehat{D})} \quad (A16)$$

and must be positive, since $f(p_i;\widehat{D})$ is negative for $0 < p_i \leq 1$.

Lastly, $(IV)$ is

$$\frac{d^2p_i}{dD_i^2} = \frac{d^2}{dD_i^2}\left[\frac{1-D_i-\frac{m}{c}+\sqrt{\left(1-D_i-\frac{m}{c}\right)^2-\frac{4ma}{c}}}{2}\right] \quad (A17)$$

$$= \frac{1}{2}\frac{-\frac{4ma}{c}}{\left(\left(1-D_i-\frac{m}{c}\right)^2-\frac{4ma}{c}\right)^{3/2}} < 0 \quad (A18)$$

(Note that the denominator of this fraction is a positive real number because $D_i < D^*$.) //

*Relationship 5: dT\*/dD<sub>i</sub> decreases without bound as D<sub>i</sub> increases toward D\*.*

This is equivalent to the statement $\lim_{D_i \nearrow D^*} \frac{dT^*}{dD_i} = -\infty$. The left side is

$$\lim_{D_i \nearrow D^*} \frac{dT^*}{dD_i} = \lim_{D_i \nearrow D^*} \left(\frac{dT^*}{dp_i}\right) \cdot \lim_{D_i \nearrow D^*} \left(\frac{dp_i}{dD_i}\right). \quad (A19)$$

Because $\frac{dT^*}{dp_i} = \frac{-1}{f(p_i(D_i); \widehat{D})}$ is a left-continuous function of $D_i$ for $D_i \leq D^*$, the first limit in the product (A19) can be found by direct substitution:

$$\lim_{D_i \nearrow D^*} \left(\frac{dT^*}{dp_i}\right) = \frac{-1}{f(p_i(D^*); \widehat{D})}. \quad (A20)$$

This limit is a finite positive number.

The second limit in the product (A19) is

$$\lim_{D_i \nearrow D^*} \left(\frac{dp_i}{dD_i}\right) = \lim_{D_i \nearrow D^*} \left(-\frac{1}{2}\left(1 + \frac{1 - D_i - m/c}{\sqrt{(1 - D_i - m/c)^2 - 4ma/c}}\right)\right)$$

$$= -\frac{1}{2}\left(1 + \lim_{D_i \nearrow D^*} \frac{\left(1 - D_i - \frac{m}{c}\right)}{\sqrt{\left(1 - D_i - \frac{m}{c}\right)^2 - \frac{4ma}{c}}}\right). \quad (A21)$$

As $D_i \nearrow D^*$, the numerator in (A21) goes to $2\sqrt{ma/c}$, a finite positive number, while the denominator approaches zero from the positive side. The limit within expression (A21) is thus $+\infty$, making expression (A21) equal to $-\infty$ and the product of the limits in (A19) equal to $-\infty$. //

*Relationship 6: dT\*/dD<sub>f</sub> < 0.* We have

$$\frac{dT^*}{dD_f} = \frac{dT^*}{dp_{thr}} \frac{dp_{thr}}{dD_f}. \quad (A22)$$

From equation (A2) and the Fundamental Theorem of Calculus, the first term in this product is

$$\frac{dT^*}{dp_{thr}} = \frac{1}{f(p_{thr}; \widehat{D})}, \quad (A23)$$

which is negative. For the second term, we differentiate $p_{thr}(D_f)$ (equation (A24)) with respect to $D_f$ to obtain

$$\frac{dp_{thr}}{dD_f} = \frac{1}{2}\left(-1 + \frac{1 - D_f - \frac{m}{c}}{\sqrt{\left(1 - D_f - \frac{m}{c}\right)^2 - \frac{4ma}{c}}}\right). \quad (A24)$$

We have that $1 - D_f - \frac{m}{c} > \sqrt{\left(1 - D_f - \frac{m}{c}\right)^2 - \frac{4ma}{c}}$, so $dp_{thr}/dD_f$ is positive. Therefore $dT^*/dD_f$ is the product of a negative and a positive term, and is negative. //

### *Relationship 7: dT\*/dD<sub>f</sub> decreases without bound as D<sub>f</sub> increases to D\*.*

This is equivalent to the statement $\lim_{D_f \nearrow D^*} \frac{dT^*}{dD_f} = -\infty$. The left side is

$$\lim_{D_f \nearrow D^*} \frac{dT^*}{dD_f} = \lim_{D_f \nearrow D^*} \left(\frac{dT^*}{dp_{thr}}\right) \cdot \lim_{D_f \nearrow D^*} \left(\frac{dp_{thr}}{dD_f}\right). \tag{A25}$$

Because $\frac{dT^*}{dp_{thr}} = \frac{1}{f(p_{thr}(D_f); \widehat{D})}$ is a left-continuous function of $D_f$ for $D_f \leq D^*$, the first limit in the product can be found by direct substitution:

$$\lim_{D_f \nearrow D^*} \left(\frac{dT^*}{dp_{thr}}\right) = \frac{1}{f(p_{thr}(D^*); \widehat{D})}. \tag{A26}$$

This is a finite negative number.

Using equation (A4), the second limit in the product is

$$\lim_{D_f \nearrow D^*} \left(\frac{dp_{thr}}{dD_f}\right) = \lim_{D_f \nearrow D^*} \left( \frac{1}{2}\left(-1 + \frac{1 - D_f - m/c}{\sqrt{(1-D_f-m/c)^2 - 4ma/c}}\right)\right) \tag{A27}$$

$$= \frac{1}{2}\left(-1 + \lim_{D_f \nearrow D^*} \frac{\left(1 - D_f - \frac{m}{c}\right)}{\sqrt{\left(1-D_f-\frac{m}{c}\right)^2 - \frac{4ma}{c}}}\right). \tag{A28}$$

As $D_f \nearrow D^*$, the numerator in this limit goes to $2\sqrt{ma/c}$, a finite positive number, while the denominator approaches zero from the positive side. The limit within expression (A28) is thus $+\infty$, making expression (A27) equal $+\infty$ and the product of the limits in (A26) equal $-\infty$. //